\documentclass[english,prb,twocolumn,showpacs]{revtex4}
\usepackage[T1]{fontenc}
\usepackage[latin1]{inputenc}
\usepackage{amsmath}
\usepackage{graphicx}
\usepackage{amssymb}

\makeatletter
\usepackage{babel}
\makeatother
\begin{document}

\title{Band convergence and linearization-error correction of all-electron
$GW$ calculations: the extreme case of zinc oxide}

\author{Christoph Friedrich, Mathias~C.~Müller, and Stefan Blügel}

\affiliation{Peter Grünberg Institut and Institute for Advanced Simulation, Forschungszentrum
Jülich and JARA, 52425 Jülich, Germany}

\begin{abstract}
Recently, Shih \emph{et al.} {[}Phys. Rev. Lett. 105, 146401 (2010){]}
published a theoretical band gap for wurtzite ZnO, calculated with
the non-selfconsistent $GW$ approximation, that agreed surprisingly
well with experiment while deviating strongly from previous studies.
They showed that a very large number of empty bands is necessary to
converge the gap. We reexamine the $GW$ calculation with the full-potential
linearized augmented-plane-wave method and find that even with 3000
bands the band gap is not completely converged. A hyperbolical fit
is used to extrapolate to infinite bands. Furthermore, we eliminate
the linearization error for high-lying states with local orbitals.
In fact, our calculated band gap is considerably larger than in previous
studies, but somewhat smaller than that of Shih \emph{et al..}
\end{abstract}


\maketitle
In a recent Letter Shih \emph{et al.} \cite{Shih} presented a new
calculation for the band gap of ZnO in the wurtzite structure employing
the $GW$ approximation \cite{Hedin} for the electronic self-energy.
They used a conventional non-selfconsistent, \emph{one-shot} approach
\cite{Hybertsen,Godby} in which neither the quasiparticle energies
nor the quasiparticle wave functions are updated. Instead, the $GW$
self-energy $\Sigma^{\sigma}(\mathbf{r},\mathbf{r}';\omega)$ {[}Eq.~(\ref{eq:Sigma}){]}
is constructed from the Kohn-Sham Green function taken from a density-functional
theory calculation with the local-density approximation (LDA) for
the exchange-correlation energy functional. The quasiparticle energy
$E_{n\mathbf{q}}^{\sigma}$ with band index $n$, Bloch vector $\mathbf{q}$,
and spin $\sigma$ is then obtained from the nonlinear equation\begin{equation}
E_{n\mathbf{q}}^{\sigma}=\epsilon_{n\mathbf{q}}^{\sigma}+\langle\varphi_{n\mathbf{q}}^{\sigma}|\Sigma^{\sigma}(E_{n\mathbf{q}}^{\sigma})-v_{\mathrm{xc}}^{\sigma}|\varphi_{n\mathbf{q}}^{\sigma}\rangle\label{eq:E}\end{equation}
with the Kohn-Sham energy $\epsilon_{n\mathbf{q}}^{\sigma}$, the
wave function $\varphi_{n\mathbf{q}}^{\sigma}(\mathbf{r})$, and the
exchange-correlation potential $v_{\mathrm{xc}}^{\sigma}(\mathbf{r})$.
Offdiagonal elements of $\Sigma^{\sigma}-v_{\mathrm{xc}}^{\sigma}$
are neglected. 

All previous calculations \cite{Usuda,Shishkin,Fuchs,Gori} invoking
the one-shot LDA+$GW$ approach showed that the band gap of wurtzite
ZnO is underestimated with respect to the experimental value by more
than 1 eV. They fall in the range 2.12--2.6~eV while the experimental
gap amounts to 3.6~eV, \cite{Tsoi} after correction for vibrational
effects. This large underestimation is untypical for $GW$ calculations
of $sp$-bound systems.

The Letter of Shih \emph{et al.}~addressed two issues: first, the
erroneous hybridization effects between Zn 3\emph{d} and O 2\emph{p}
states that results from the self-interaction error within the LDA,\cite{Perdew}
and second, the band convergence in the correlation part of the self-energy.
The first problem was tackled with the LDA+$U$ approach, \cite{Anisimov}
in which an orbital-dependent potential corrects the position of the
3\emph{d} bands and, thus, reduces hybridization effects with the
O 2\emph{p} states. However, the combination LDA+$U$ and $GW$ yields
a band gap that is still well below the experimental value. Therefore,
the authors investigated the second issue by carefully converging
the correlation self-energy and the dielectric matrix with respect
to the number of bands. They performed calculations with up to 3000
bands corresponding to a maximal band energy of 67 Ry as well as a
cutoff for the dielectric matrix of up to 80 Ry and showed that the
resulting $GW$ band gaps, 3.4~eV for LDA+$GW$ and 3.6~eV for LDA+$U$+$GW$,
turned out to be in very good agreement with experiment. They also
demonstrated that a too small energy cutoff for the dielectric matrix
can lead to a false convergence behavior: the band gap seems to converge
with many fewer bands, but toward a value that is too small.

These new results for ZnO are in striking contrast to previous studies.
If they are correct, they cast doubt on all $GW$ calculations published
so far, not only for ZnO but also for other materials, especially
for systems with localized states. In fact, Shih \emph{et al.} point
out at the end of their paper that {}``many of the previous quasiparticle
calculations ... involving localized states may need to be reexamined.''
This will likely provoke a controversial debate that reaches different
fields of solid state physics and requests a rapid clarification. 

Since its publication the paper has aroused criticism, mostly from
the all-electron community who pointed out that calculations with
the pseudopotential approximation cannot give a definite answer for
the real $GW$ gap because of the neglect of the core-valence exchange
effects, the pseudized form of the valence wave functions, and the
inaccurate description of high-lying states. Therefore, they attributed
the large difference between the new result and the previous all-electron
calculations\cite{Fuchs,Usuda,Shishkin} (2.12--2.44~eV) to the approximations
inherent to the pseudopotential approach or to numerical errors in
the calculation. It is, thus, vitally important that the $GW$ band
gap of ZnO is reinvestigated and thoroughly converged with a genuine
all-electron method to provide a standard for the discussion that
will follow.

In this paper, we present an all-electron LDA+$GW$ calculation for
wurtzite ZnO that is based on the full-potential linearized augmented-plane-wave
(FLAPW) method.\cite{Weinert} For simplicity, we restrict ourselves
to the standard LDA approach for the noninteracting starting point
without an additional Hubbard $U$ parameter. Our calculation yields
a band gap that is, in fact, much larger than that of the previous
calculations, but somewhat smaller than the result of Shih \emph{et
al..} We go beyond their approach in two respects: we employ neither
the pseudopotential approximation nor a plasmon-pole model for the
dielectric matrix. Instead, the screened interaction\begin{eqnarray}
W(\mathbf{r},\mathbf{r}';\omega) & = & v(\mathbf{r},\mathbf{r}')+\iint v(\mathbf{r},\mathbf{r}'')P(\mathbf{r}'',\mathbf{r}''';\omega)\nonumber \\
 &  & \times W(\mathbf{r}''',\mathbf{r}';\omega)\, d^{3}r''d^{3}r'''\label{eq:W}\end{eqnarray}
 is calculated explicitly within the random-phase approximation for
the polarization function\begin{eqnarray}
P(\mathbf{r},\mathbf{r}';\omega) & = & -\frac{i}{2\pi}\sum_{\sigma}\int_{-\infty}^{\infty}G^{\sigma}\left(\mathbf{r},\mathbf{r}',\omega'\right)\nonumber \\
 &  & \times G^{\sigma}\left(\mathbf{r}',\mathbf{r},\omega'-\omega\right)d\omega'\,,\end{eqnarray}
where the Green function $G^{\sigma}\left(\mathbf{r},\mathbf{r}',\omega\right)$
is constructed from Kohn-Sham energies and wave functions. The frequency
convolution of the self-energy\begin{equation}
\Sigma^{\sigma}(\mathbf{r},\mathbf{r}';\omega)=\frac{i}{2\pi}\int_{-\infty}^{\infty}G^{\sigma}(\mathbf{r},\mathbf{r}';\omega+\omega')W(\mathbf{r},\mathbf{r}';\omega')e^{i\eta\omega'}\, d\omega'\label{eq:Sigma}\end{equation}
($\eta$ is a positive infinitesimal) is evaluated analytically for
$v(\mathbf{r},\mathbf{r}')$ {[}see Eq.~(\ref{eq:W}){]}, and with
a contour integration \cite{Godby,Aryasetiawan} on the complex frequency
plane for the remainder $W(\mathbf{r},\mathbf{r}';\omega)-v(\mathbf{r},\mathbf{r}')$.
The nonlinear equation (\ref{eq:E}) is solved on an energy mesh around
$\epsilon_{n\mathbf{q}}^{\sigma}$ with a cubic spline interpolation
between the mesh points. Thus, no additional Taylor expansion of the
self-energy is needed. Details of the implementation can be found
in Ref.~\onlinecite{Friedrich}.

We carefully converged the number of empty bands for the calculation
of both the polarization function and the correlation self-energy
as well as the parameters for the all-electron mixed product basis,\cite{Kotani,Friedrich}
in which we represent the dielectric matrix. While the ground-state
electron density was converged with a standard LAPW basis with moderate
cutoff parameters ($l_{\mathrm{max}}=8$, $G_{\mathrm{max}}=4.3\, a_{0}^{-1}$,
where $a_{0}$ is the Bohr radius), we had to employ much larger cutoffs
to generate enough wave functions for the $GW$ calculation: a linear
momentum cutoff of $G_{\mathrm{max}}=8.0\, a_{0}^{-1}$ and an angular
momentum cutoff in the muffin-tin (MT) spheres of $l_{\mathrm{max}}=12$.
Furthermore, in order to avoid linearization errors in the MT part
of the LAPW basis,\cite{Friedrich2006,Krasovskii} we added local
orbitals \cite{Singh,Sjoestedt} (LOs) with different angular momentum
quantum numbers and energy parameters distributed over the relevant
energy range: 292 LOs for Zn (five LOs for each $lm$ channel with
$l=0$-$3$, three for $l=4$, two for $l=5$, and one for $l=6$)
and 186 for O (four LOs for $l=0$-$3$, two for $l=4$, and one for
$l=5$). We also treat the $3s$ and $3p$ semicore states of Zn explicitly
with LOs.

For the mixed product basis we found an angular momentum cutoff of
$4$ in the spheres and a suprisingly small linear momentum cutoff
of 2.4~$a_{0}^{-1}$ in the interstitial region to be sufficient.
However, we had to take into account many product functions in the
MT spheres, which after optimization led to 177 MT functions for Zn
(ten, eight, eight, seven, and six radial functions per $lm$ channel
for $l=0$-$4$, respectively) and 127 for O (eight, six, six, five,
and four radial functions for $l=0$-$4$, respectively). Obviously,
the rapid variations close to the atomic nuclei must be accurately
described. Within the all-electron mixed product basis this is possible
with a relatively modest number of MT functions, while in a pure plane-wave
approach a very large number of plane waves is necessary to resolve
the variations adequately. This explains the finding of Shih \emph{et
al.} that the dielectric matrix must be converged to very large energy
cutoffs. The total number of mixed product basis functions in the
calculations is less than 700 per $\mathbf{k}$ point. This number
is further reduced to around 490 by constructing linear combinations
that are continuous in value and radial derivative at the MT sphere
boundaries.\cite{Betzinger}

\begin{figure}
\includegraphics[%
  width=0.70\columnwidth,
  angle=-90]{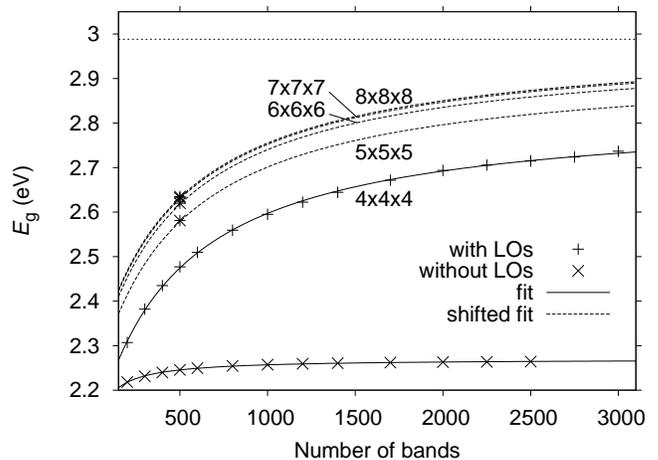}

\caption{\label{cap:bandconv}Band convergence of the quasiparticle band gap
of ZnO employing a 4$\times$4$\times$4 $\mathbf{k}$-point set and
calculated with (pluses) and without local orbitals (LOs) (crosses)
for high-lying states. The solid lines show the hyperbolical fits.
We also indicate results with finer $\mathbf{k}$-point samplings
(stars) calculated with LOs and 500 bands. The dashed lines show the
hyperbolical fit shifted to align with these results. The fit asymptote
for the 8$\times$8$\times$8 $\mathbf{k}$-point set at 2.99~eV
(dotted line) is considered the best estimate for the all-electron
one-shot $GW$ band gap.}
\end{figure}
Figure \ref{cap:bandconv} shows the quasiparticle band gap of ZnO
as a function of the number of states included in the calculation
of the polarization function and the correlation self-energy. The
calculations were performed with a 4$\times$4$\times$4 $\mathbf{k}$-point
sampling of the Brillouin zone. There is a large difference between
calculations with (pluses) and without the LOs for unoccupied states
(crosses), which shows the importance of eliminating the linearization
error of the conventional LAPW basis. As the linearization error becomes
larger for higher and higher bands, it is not surprising that the
difference between the convergence curves grows toward increasing
numbers of bands. We find an asymptotic difference of 0.5~eV. The
calculations without LOs for unoccupied states already converge with
a few hundred bands, which could have led the authors of the previous
all-electron studies to believe that their calculations are sufficiently
converged. In fact, when the converged value of 2.27~eV is corrected
with respect to finer $\mathbf{k}$-point samplings, we arrive at
2.44~eV, which lies at the upper edge of the range of the all-electron
$GW$ band gaps published so far.

\begin{figure}
\includegraphics[%
  width=0.70\columnwidth,
  angle=-90]{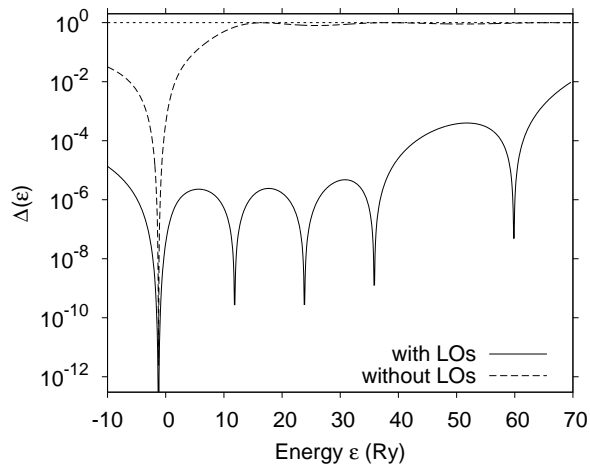}

\caption{\label{cap:basis}Representation error $\Delta(\epsilon)$ (see text)
of the set of spherical MT functions at an oxygen atom with (solid
line) and without local orbitals (LOs) (dashed line) as a function
of energy. The Fermi energy is set to $\epsilon=0$. The {}``spikes''
clearly identify the positions of the energy parameters. In the case
without LOs, the representation error soon approaches unity (dotted
line) for energies above 10~Ry, which means that the basis becomes
practically orthogonal to the exact solution of the radial scalar-relativistic
Dirac equation.}
\end{figure}
The linear momentum cutoff allows the interstitial part of the LAPW
basis to be converged in a systematic way. In the MT spheres, however,
the basis is linearized around predefined energy parameters, which
gives rise to the linearization error for high-lying states. We now
analyze this error in more detail and show that it can be eliminated
very effectively with the LOs. Figure \ref{cap:basis} shows the representation
error $\Delta(\epsilon)$ of the MT basis at an oxygen atom for the
angular momentum $l=0$ as a function of energy. We define\begin{equation}
\Delta(\epsilon)=\int_{0}^{S}r^{2}[R(\epsilon,r)-R_{\mathrm{r}}(\epsilon,r)]^{2}dr\,,\end{equation}
where $S$ is the oxygen MT sphere radius, $R(\epsilon,r)$ is the
normalized solution of the radial scalar-relativistic Dirac equation
(cf.~Ref.~\onlinecite{Friedrich2006}) for the energy $\epsilon$
and the angular momentum $l=0$, and $R_{\mathrm{r}}(\epsilon,r)$
is its best representation in terms of linear combination of the MT
functions contained in the LAPW basis. In the conventional basis there
are only two radial functions available: the solution $R(\epsilon_{\mathrm{par}},r)$
of the radial Dirac equation for $l=0$ at the energy parameter $\epsilon_{\mathrm{par}}$,
in this case $\epsilon_{\mathrm{par}}=-1.24$~Ry, and its energy
derivative $dR(\epsilon_{\mathrm{par}},r)/d\epsilon_{\mathrm{par}}$.
{[}In the literature these functions are commonly denoted by $u_{l=0}(r)$
and $\dot{u}_{l=0}(r)$, respectively.{]} The dashed line in Fig.~\ref{cap:basis}
shows that this conventional basis represents the occupied states
very accurately, which are located in the small energy range $-1.33\le\epsilon\le0\,\textrm{Ry}$,
but fails to describe unoccupied states at higher energies. In fact,
at energies above 10~Ry the exact solution $R(\epsilon,r)$ becomes
practically orthogonal to the MT basis, and the representation error
approaches unity. Interestingly, at this energy, which roughly corresponds
to the 500th band, the convergence curve for the calculations without
the LOs levels off, and the band gap seemingly converges (see Fig.~\ref{cap:bandconv}).
The description of wave functions at higher energies is considerably
improved, if LOs are added to the LAPW basis. The solid line shows
the corresponding representation error with four additional LOs at
12, 24, 36, and 60~Ry. As can be seen, the error remains below $10^{-3}$
for a very large energy range up to 65~Ry. 

As Fig.~\ref{cap:bandconv} shows, the calculations with eliminated
linearization error (pluses) take far more bands to converge. In fact,
even with 3000 bands the band gap is still not completely converged.
Therefore, we fitted the values with the hyperbolical function \begin{equation}
f(N)=\frac{a}{N-N_{0}}+b\end{equation}
 where $a$, $b$, and $N_{0}$ are fit parameters. It is surprising
how closely the fitted curve (solid line) follows the calculated data
points. This makes us confident in taking the fit asymptote $b$ as
the band gap extrapolated to infinite bands. Furthermore, we have
recalculated the band gap with finer $\mathbf{k}$-point meshes (up
to 8$\times$8$\times$8) and 500 bands (crosses). The dashed lines
show correspondingly shifted hyperbolical fits. The asymptote of the
fit corresponding to an 8$\times$8$\times$8 $\mathbf{k}$-point
sampling is found at 2.99~eV, which we take as the final best estimate
for the all-electron one-shot $GW$ band gap.

This band gap is 0.4--0.9~eV larger than the previously reported
values. Both the large number of bands that are needed for proper
convergence and the elimination of the linearization error, which
has not been undertaken in the previous all-electron studies, are
responsible for this large difference. Our value is still about 0.4~eV
smaller than the band gap of Shih \emph{et al.}, though. In fact,
a certain discrepancy should be expected because of the pseudopotential
approximation and the plasmon-pole model for the dielectric function
used in Ref.~\onlinecite{Shih}. The pseudopotential approximation
not only neglects the important contribution of core-valence exchange.
It also yields accurate wave functions only in the vicinity of the
atomic electron energies, but not for high-lying states. This error
is very similar in spirit to the linearization error of the LAPW basis
and is also of the same magnitude.\cite{Friedrich2006} Thus, it should
have an impact on the $GW$ results comparable in size to the linearization
error, whose elimination gives rise to a sizable correction of 0.5~eV,
as we have shown in the present work.

With the LDA band gap of only 0.84~eV the quasiparticle correction
amounts to more than 2~eV. It can be expected that a treatment beyond
the one-shot approach, for example, by taking into account offdiagonal
elements of the self-energy, by updating the Green function, or by
including vertex corrections, will further increase the value and,
thus, bring it even closer to the experimental value. As was shown
in Ref.~\onlinecite{Shih}, already using LDA+$U$ instead of LDA
as the mean-field starting point, which corrects the 2$p$-3$d$ hybridization,
gives an upward correction of 0.2~eV in the resulting $GW$ band
gap.

In conclusion, the band convergence is a serious issue in $GW$ calculations
and must be thoroughly dealt with. ZnO is an extreme case in this
respect. We have reexamined the calculation of the ZnO band gap by
Shih \emph{et al.},\cite{Shih} and could confirm their main result:
the $GW$ band gap of ZnO shows a very slow convergence with respect
to the number of states used to construct the polarization function
and the correlation self-energy. Furthermore, when high-lying bands
are included in the calculation, the linearization error of all-electron
approaches becomes another important issue. As we have shown, it can
be eliminated systematically within the FLAPW method by augmenting
the basis in the MT spheres with LOs. In the case of ZnO this procedure
yields a correction of about 0.5~eV and brings the calculated band
gap (2.99~eV) much closer to experiment than in previous studies.
We believe that our study helps to clarify the contradiction between
the pseudopotential results of Ref.~\onlinecite{Shih} and the previous
results based on all-electron approaches. \cite{Usuda,Shishkin,Fuchs}

We acknowledge helpful discussions with Andreas Gierlich and Georg
Kresse. This work was supported in part by the Deutsche Forschungsgemeinschaft
through the Priority Program 1145.

\end{document}